\documentstyle[epsfig]{aipproc}
\begin{document}
\title{Gravitational waves from inspiral into massive black holes}
\author{Scott A.\ Hughes}
\address{
Theoretical Astrophysics, California Institute of Technology,
Pasadena, CA 91125}
\maketitle
\begin{abstract}
Space-based gravitational-wave interferometers such as LISA will be
sensitive to the inspiral of stellar mass compact objects into black
holes with masses in the range of roughly $10^5$ solar masses to (a
few) $10^7$ solar masses.  During the last year of inspiral, the
compact body spends several hundred thousand orbits spiraling from
several Schwarzschild radii to the last stable orbit.  The
gravitational waves emitted from these orbits probe the strong-field
region of the black hole spacetime and can make possible high
precision tests and measurements of the black hole's properties.
Measuring such waves will require a good theoretical understanding of
the waves' properties, which in turn requires a good understanding of
strong-field radiation reaction and of properties of the black hole's
astrophysical environment which could complicate waveform generation.
In these proceedings, I review estimates of the rate at which such
inspirals occur in the universe, and discuss what is being done and
what must be done further in order to calculate the inspiral waveform.
\end{abstract}

One of the most exciting sources that should be measured by the
space-based gravitational-wave interferometer LISA (the Laser
Interferometer Space Antenna {\cite{sah:lisa}}) is the inspiral of a
``small'' ($1-10\,M_\odot$) compact body into a massive
($10^{5-7}\,M_\odot$) black hole.  Such massive black holes reside at
the cores of galaxies; the smaller compact bodies will become bound to
the hole and spiral into them after undergoing interactions with stars
and other objects in the environment of the galactic center.  The
measurement of gravitational waves from such inspirals will make
possible very high precision tests of general relativity, and probe
the nature of the galactic core's environment.

To set the stage for understanding why these inspirals are such
interesting objects, consider the following estimates: the orbital
energy of a small body in an equatorial, prograde orbit of a Kerr
black hole is
\begin{equation}
E^{\rm orb} = \mu{{1 - 2 v^2 + q v^3}\over\sqrt{1 - 3 v^2 + 2 q v^3}}\;,
\label{eq:sah:kerrorbenergy}
\end{equation}
where $v \equiv \sqrt{M/r}$ and $q \equiv a/M$.  (I use units in which
$G = c = 1$ throughout.)  The orbital frequency of the small body is
\begin{equation}
\Omega = {M^{1/2}\over{r^{3/2} + a M^{1/2}}}\;.
\label{eq:sah:kerrorbfreq}
\end{equation}
Radiation reaction carries orbital energy away from the system,
causing the orbit to shrink.  Eventually it shrinks enough that the
body reaches the innermost stable circular orbit (ISCO).  Orbits
inside this radius are dynamically unstable; further radiative
evolution tends to push the body into the black hole.

Post-Newtonian theory allows us estimate the rate at which the system
loses energy as a power series in the quantity $u \equiv
(M\Omega)^{1/3}$ (which is roughly the orbital speed) and the black
hole's spin $a$.  Reference {\cite{sah:minoetal97}} gives the energy
loss in such a post-Newtonian expansion:
\begin{equation}
{dE\over dt} = -{32\over 5}\left({\mu\over M}\right)^2 u^{10}
\left[f_{\rm Schw.}(u) + f_{\rm spin}(a,u)\right]\;.
\label{eq:sah:dEdt}
\end{equation}
The prefactor $-32/5 (\mu/M)^2 u^{10}$ is the result one gets applying
the quadrupole formula to a binary described with Newtonian gravity;
the function $f_{\rm Schw.}(u)$ is a (rather high-order) correction
appropriate for zero-spin black holes, and $f_{\rm spin}(a,u)$ is a
correction incorporating information about the hole's spin.  (Note
that this formula is only appropriate for $\mu \ll M$: it does {\it
not} incorporate any finite mass ratio corrections.)

Equations (\ref{eq:sah:kerrorbenergy})--(\ref{eq:sah:dEdt}) can
be used to estimate the time it takes for a small body to spiral
from radius $r_1$ to $r_2$, and the number of gravitational-wave
cycles it emits in that time:
\begin{eqnarray}
T &=& \int_{t_1}^{t_2} dt = \int_{r_1}^{r_2} {dt\over dr}\,dr
  = \int_{r_1}^{r_2} {dE/dr\over dE/dt}\,dr\;,\\
\label{eq:sah:inspiraltime}
N_{\rm cyc} &=& \int_{t_1}^{t_2} f_{\rm gw}\,dt
= \int_{r_1}^{r_2} {\Omega\over\pi} {dt\over dr}\,dr
= \int_{r_1}^{r_2} {\Omega\over\pi} {dE/dr\over dE/dt}\,dr\;.
\label{eq:sah:inspiralcycs}
\end{eqnarray}
(On the last line, I have assumed that the bulk of the radiation comes
out in the quadrupole $m = 2$ mode, so that $f_{\rm gw} = 2\Omega/2
\pi$.)  Consider now the inspiral of a $5\,M_\odot$ body into a
rapidly spinning ($a \simeq M$) $10^6\,M_\odot$ black hole.  The small
body spirals from a radius of $8 M$ (in Boyer-Lindquist coordinates)
to the ISCO in one year, emitting around $5 \times 10^5$
gravitational-wave cycles as it does so.  The gravitational-waves that
it emits lie in the frequency band $3\times10^{-3}\,\mbox{Hz} \lesssim
f \lesssim 3\times10^{-2}\,\mbox{Hz}$ --- the band to which LISA is
most sensitive.  Indeed, careful analyses of the detectability of the
signal by LISA {\cite{sah:finnthorne}} indicate that such an inspiral
should be detected out to a distance of roughly 1 Gigaparsec with
amplitude signal-to-noise ratio of around $10$ to $100$ (depending on
factors such as the mass of the small body, the mass of the black
hole, and the black hole's spin).  The fact that such a large number
of cycles are emitted indicates that details of the waveform can in
principle be measured to very high precision.  {\it Detection of
extreme mass-ratio inspirals by LISA offers the possibility of very
high precision measurements of the characteristics of extreme
strong-field regions of spacetime.}

The high precision tests that LISA will be able to make should allow
us to directly map the characteristics of the massive body's spacetime
metric and confirm that they in fact exhibit the Kerr metric.  Most
likely, the way that this will be done will be to measure the
multipole moments of the massive body.  Fintan Ryan {\cite{sah:ryan}}
has shown that the gravitational waves which are emitted as a small
body spirals into a massive compact object contain a ``map'' of the
massive body's spacetime.  By measuring the gravitational waves and
decoding the map, one learns the mass and current multipole moments
which characterize the massive body.  All multipole moments of Kerr
black holes are parameterized by the holes' mass $M$ and spin $a$:
\begin{equation}
M_l + i S_l = M\,(i a)^l\;.
\label{eq:sah:kerrmoments}
\end{equation}
For Kerr black holes, knowledge of the moments $M_0 \equiv M$ and $S_1
\equiv aM$ determines {\it all} higher moments.  This is one way of
stating the ``no-hair'' theorem: The macroscopic properties of a black
hole are entirely determined by its mass and spin.  (I neglect the
astrophysically uninteresting possibility of charged holes.)  {\it By
measuring gravitational waves from extreme mass ratio inspiral and
thereby mapping the massive body's spacetime, LISA will test the
no-hair theorem for black holes, determining whether the massive body
has multipole moments characteristic of the Kerr metric, or whether
the body is something more exotic, such as a boson star.}

The waves emitted by extreme mass ratio inspiral are thus likely to be
directly measureable by LISA, and are likely to be extremely
interesting.  One might next wonder whether they occur often enough to
be interesting.  This question has been examined in some detail by
Martin Rees and Steinn Sigurdsson {\cite{sah:rs,sah:sig}}.  They
consider the scatter of stellar mass black holes in the central
density cusp of galaxies into tightly bound orbits of the galaxies'
central black hole.  Occasionally, such a scattering event will put
the stellar mass hole into an orbit which is so tightly bound that its
future dynamics are driven by gravitational-wave emission, and it
becomes an interesting source for LISA.  They find that the rate of
such events is likely to lie in the range
\begin{equation}
{\mbox{1 event}\over\mbox{year }\mbox{Gpc}^3} \lesssim
{\cal R}\lesssim{\mbox{1 event}\over\mbox{month }\mbox{Gpc}^3}\;.
\label{eq:sah:rate}
\end{equation}
Obviously, there are large uncertainties in this calculation.  The low
mass end of the massive black hole population (which is most relevant
to LISA observations) is not as well constrained as the population of
very massive black holes ($M\gtrsim10^8\,M_\odot$), and there are
uncertainties in the rate at which stellar mass black holes are
``fed'' into the central hole to produce extreme mass ratio binaries.
However, the lower end of the rate estimate ({\ref{eq:sah:rate}}) is
based on very conservative estimates.  We may rather robustly estimate
that the rate measured by LISA will be several events per year out to
a Gigaparsec {\cite{sah:steinn_pc}}.

The waves that LISA will measure from these sources will come from
orbits that are rather eccentric and inclined with respect to the
black hole's equatorial plane {\cite{sah:hilsbender}}.  To best
interpret the measured waves (and, indeed, in order to improve the
odds of seeing the waves at all in the detector's noisy data stream),
it will be necessary to have some theoretical modeling of the orbit
and the waves that it emits as gravitational radiation reduces the
orbit's energy and angular momentum.  We expect that the radius,
inclination angle, and eccentricity of the orbit will change as
radiative backreaction drives the system's evolution.  Some means of
understanding these changes, in detail, is needed in order to model
the waves' evolution accurately.

Before discussing work in radiation reaction, it is necessary to take
a sanity check.  Relativity theorists often work in a very idealized
universe: their extreme mass ratio binary is likely to consist of a
big black hole, a small body, and gravitational waves.  In the real
astrophysical world, there will be complications to this pretty (but
highly idealized) picture.  One should worry whether the complications
render the relativity theorist's modeling invalid.

Perhaps the most important such complication arises from interaction
between the small inspiraling body and material accreting onto the
massive black hole.  Recently, Narayan has analyzed this interaction
and concluded that, in almost all cases, accretion induced drag is
unlikely to significantly influence extreme mass ratio inspiral
{\cite{sah:narayan}}.  This conclusion is based on the fact that in
the majority of cases, the rate at which the central black hole
accretes gas from its environment is rather low (several orders of
magnitude less than the Eddington rate {\cite{sah:adaf}}).  For these
``normal'' galaxies, much evidence {\cite{sah:narayan,sah:adaf}}
suggests that the gas accretes via an advection dominated accretion
flow (ADAF).  Narayan's calculation {\cite{sah:narayan}} shows that
the timescale for ADAF drag to change the orbit's characteristics
({\it e.g.}, the orbital angular momentum) is many (9 to 16) orders of
magnitude longer than the timescale for radiation reaction to change
the orbit's characteristics.  Thus, the relativity theorist's
idealized view of an extreme mass ratio binary is probably quite
accurate: radiation reaction is likely the most important factor
driving the evolution of extreme mass ratio binaries.

When the mass ratio is extreme, one can analyze the spacetime of the
binary using a perturbative expansion: the spacetime metric can be
written as a ``background'' from the central object (which I will
assume from now on is a Kerr black hole), plus a perturbation due to
the inspiraling body:
\begin{equation}
g_{\alpha\beta} = g^{\rm Kerr}_{\alpha\beta}(M,a) +
h_{\alpha\beta}(\mu)\;.
\label{eq:sah:perturbsplit}
\end{equation}
The evolution of the perturbation $h_{\alpha\beta}(\mu)$ should then
describe the dynamical evolution of the system.  To linear order in
the mass ratio $\mu/M$ (which should be adequate for extreme mass
ratios), this evolution can be described using perturbation
techniques, such as the Teukolsky equation\footnote{The Teukolsky
equation actually describes the evolution of a curvature quantity
related to the perturbation.} {\cite{sah:teuk72}}.

When the mass ratio is extreme, the effects of radiation reaction are
gentle enough that the system's evolution is adiabatic: the radiation
reaction timescale is much longer than the orbital timescale.  At any
given moment, the trajectory of the small body is very nearly a
geodesic, parameterized by the three constants of Kerr orbital motion:
the energy $E$, the ($z$-component of) angular momentum $L_z$, and the
``Carter constant'' $Q$.  Gravitational-wave emission causes these
three constants to change on the radiation reaction timescale.  In
this adiabatic limit, the evolution of the system can be understood in
terms of the evolution of the quantities $(E,L_z,Q)$.

It is well known that gravitational waves carry energy and angular
momentum.  One might think that they carry ``Carter constant'' as
well, and that therefore one might be able to deduce the effects of
radiation reaction by measuring the flux of radiation at infinity and
down the event horizon.  By measuring the amount of $E$, $L_z$, and
$Q$ carried in the flux one should be able to deduce how much $E$,
$L_z$, and $Q$ are lost from the orbit.  This would then allow one to
figure out orbits of Kerr black holes radiatively evolve.

This approach does not work.  One can deduce the change in the orbit's
$E$ and $L_z$ by examining radiation fluxes, but one cannot so deduce
the change in $Q$:
\begin{eqnarray}
\delta E_{\rm orbit} &=& -\delta E_{\rm radiated}\;,\nonumber\\
\delta L_{z,{\rm orbit}} &=& -\delta L_{z,{\rm radiated}}\;,\nonumber\\
\delta Q_{\rm orbit} &\ne& -\delta Q_{\rm radiated}\;.
\label{eq:sah:orbitchange}
\end{eqnarray}
The change $\delta Q$ turns out to depend explicitly on the local
radiation reaction force, $f^\mu = dp^\mu/d\tau$, which the small body
experiences due to radiative backreaction (see {\cite{sah:paperI}}).
The properties of this force (and programs to calculate it) are
described elsewhere in this volume {\cite{sah:mino_burko}}.  Here, it
is sufficient to note that an understanding of the radiation reaction
force for gravitational radiation reaction lies some time in the
future, so that we cannot yet evolve generic Kerr black hole orbits.

There are special cases where evolution of the Carter constant is not
such a nasty impediment.  One case is the evolution of equatorial
orbits.  Equatorial orbits have $Q = 0$; and, one can easily show that
an orbit which starts off equatorial remains equatorial.  In this
case, one need only evolve the energy and angular momentum.  The local
radiation reaction force is not needed in this case.  Another case is
the evolution of circular, non-equatorial orbits.  (For non-zero spin,
``circular'' means ``constant Boyer-Lindquist coordinate radius''.)
Such orbits have recently been shown to remain circular under
adiabatic radiation reaction {\cite{sah:circulartheorems}}.  Thus, in
an adiabatic evolution, a system which is initially circular and
inclined will remain circular and inclined: the system evolves through
a sequence of orbits changing only its radius and inclination angle.
By imposing ``circular goes to circular'', one can write down a simple
rule relating the change of the Carter constant to the flux of energy
and angular momentum: ${\dot Q} = {\dot Q}({\dot E}, {\dot L_z})$.

\begin{figure}
\centerline{
\epsfig{file=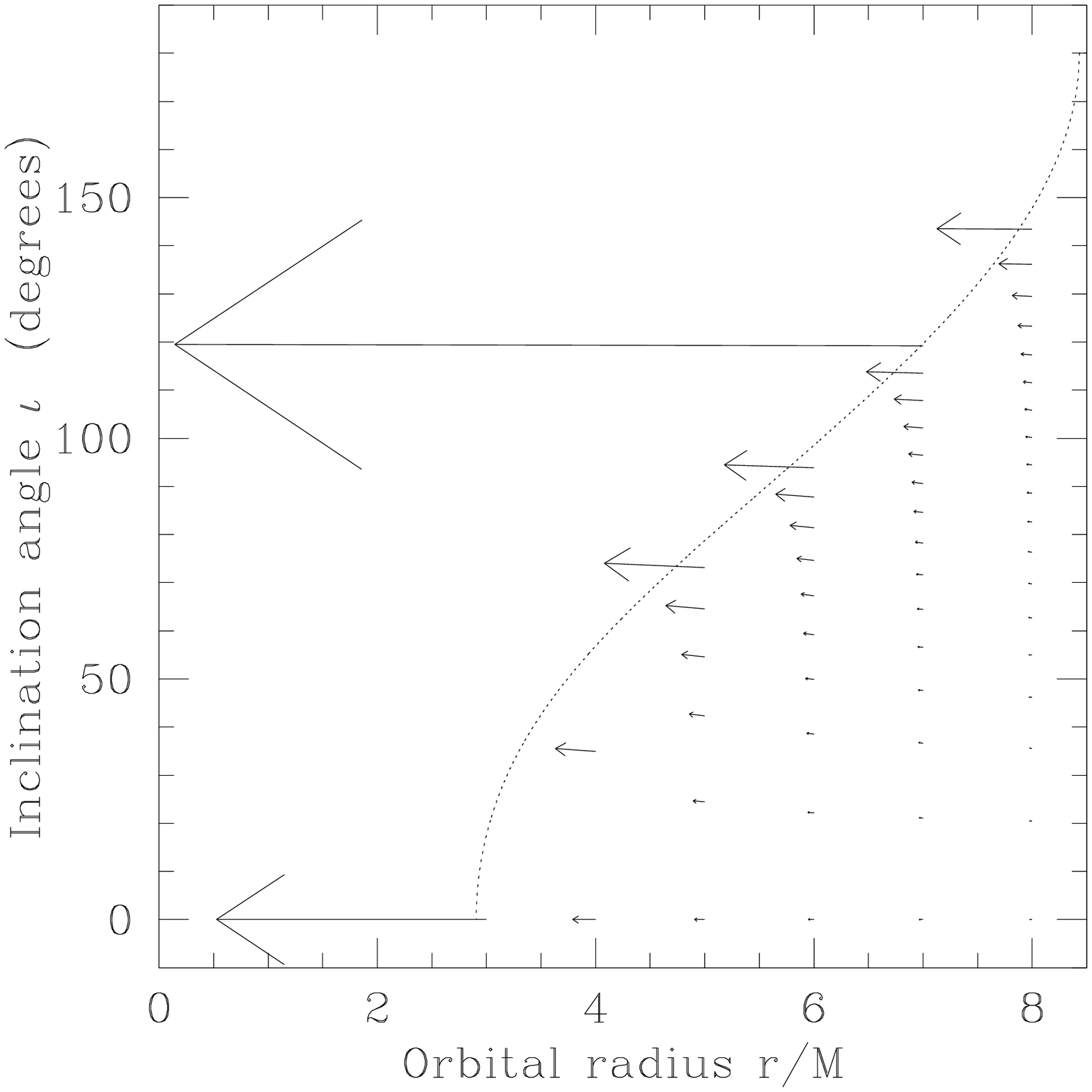,width=7cm}
\epsfig{file=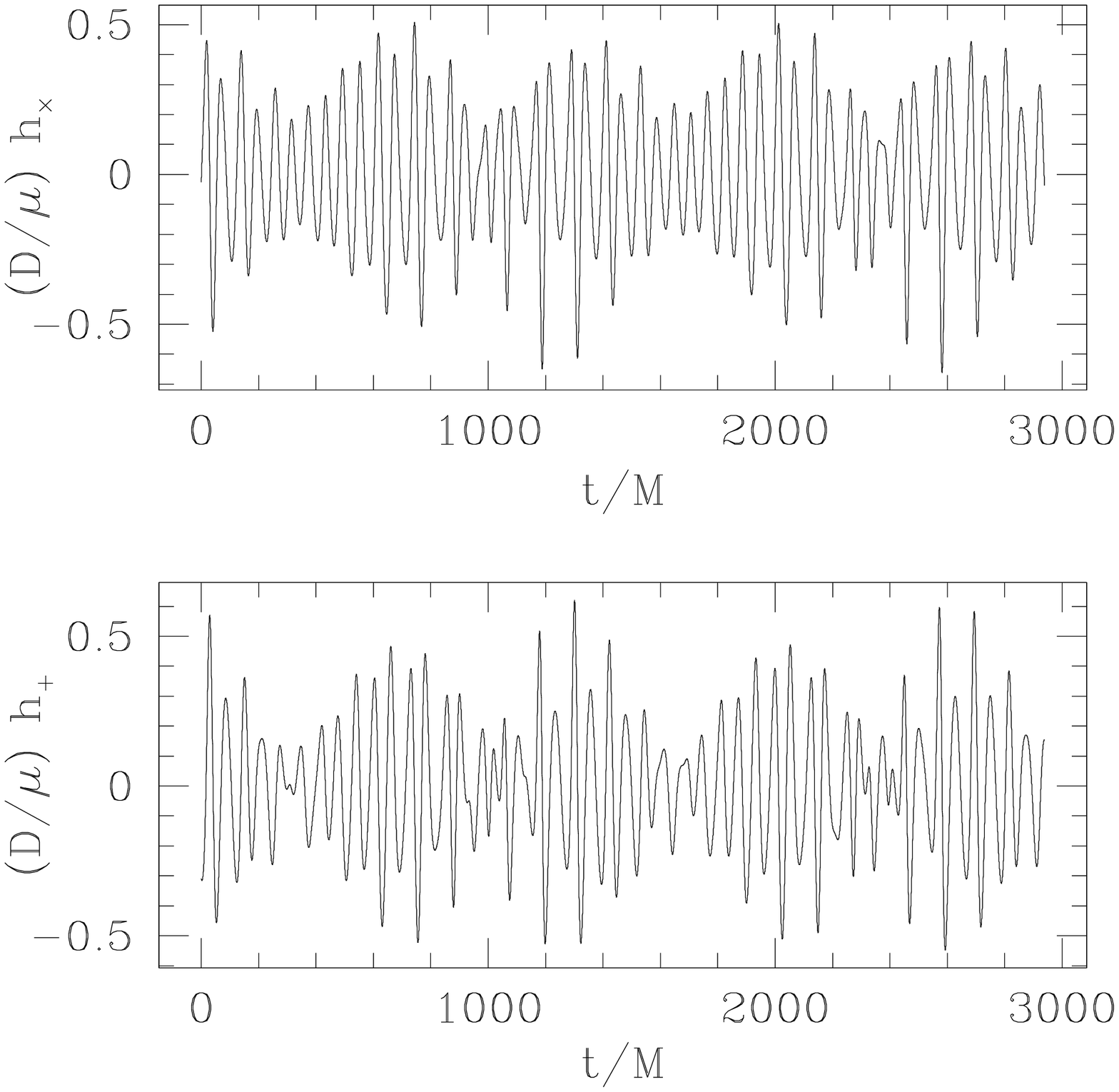,width=7cm}
}
\vskip 0.2cm
\caption{\label{fig:sah:radreact}
The left panel illustrates the effect of radiation reaction on
circular, inclined orbits.  It is for a black hole with $a = 0.8 M$.
The tail of each arrow represents a particular orbit with radius $r$
and inclination angle $\iota$.  The arrow indicates the direction in
which radiation reaction drives the orbit; the size of the arrow
indicates how quickly it is so driven.  The right panel is a snapshot
of the gravitational waveform emitted during an inspiral into a hole
with $a = 0.95 M$.  At this moment, the small body is spiraling
through $r = 7M$, $\iota = 62.43^\circ$.}
\end{figure}

Recently, I have examined the evolution of these circular orbits under
adiabatic radiation reaction, using a flux-measuring formalism based
on the Teukolsky equation.  The formalism and results are presented at
length in {\cite{sah:paperI}}.  Some highlights of the results are
presented in Figure {\ref{fig:sah:radreact}}.

Consider the left panel of Figure {\ref{fig:sah:radreact}}.  The
horizontal axis is orbital radius $r$; the vertical axis is
inclination angle $\iota$.  This figure shows the direction, in
$(r,\iota)$ phase space, in which radiation reaction tends to push the
orbit.  This particular analysis is for a black hole with $a = 0.8 M$.
Notice that the orbits are in the strong-field of the hole; the dotted
line indicates the maximum inclination angle which the orbit can have
and remain stable.  Orbits tilted beyond this line are dynamically
unstable to small perturbations and plunge into the hole.  The tail of
each arrow represents a particular orbit.  The direction of the arrow
gives the direction in which that orbit tends to evolve due to
gravitational-wave emission; the arrow's length indicates the relative
rate of evolution.  In all cases, the direction of the arrow is such
that the inclination angle {\it increases}: radiation reaction tends
to make tilted orbits more inclined.  This is exactly what one would
have predicted by extrapolating from post-Newtonian theory
{\cite{sah:ryan_pn}}.  The rate at which this inclination angle
increases is rather slow --- the aspect of the arrows in Figure
{\ref{fig:sah:radreact}} is nearly flat.  Indeed, the value of
$d\iota/dt$ in this strong-field regime is roughly 3 times smaller
than what post-Newtonian theory predicts\footnote{Recent analyses are
showing that in the extreme strong-field of rapidly rotating holes,
the inclination angle changes rather more dramatically, and in the
opposite direction: radiation reaction tends to {\it decrease} the
inclination angle.  This work is in progress {\cite{sah:paperII}}.}.
An interesting feature of Figure {\ref{fig:sah:radreact}} is the very
long arrow at $r = 7 M$, $\iota\simeq 120^\circ$.  This orbit lies
extremely close to the marginally stable orbit: it is at $\iota =
119.194^\circ$; the marginally stable orbit is at $\iota =
119.670^\circ$.  This orbit is barely dynamically stable, so a small
push has drastic effects.

The right panel of Figure {\ref{fig:sah:radreact}} shows a portion of
the gravitational waveform emitted during inspiral.  The central black
hole in this case has $a = 0.95 M$; it lies at luminosity distance $D$
from the detector.  The waveform here is shown as the small body
passes through $r = 7 M$, $\iota = 62.43^\circ$.  Note the
low-frequency modulation of both polarizations.  This is due to the
frame dragging induced precession of the orbital plane ---
Lense-Thirring precession.  Note also the many sharp, short-timescale
features present in the two polarizations.  When the spin is high,
many harmonics of the small body's fundamental orbital frequencies
contribute to the gravitational waveform.  This leads to a rather
complicated structure; the energy spectrum corresponding to these
waveforms extends to rather high frequencies.  Accurate measurement of
such complicated waveforms will be quite a challenge.  However, the
payoff is likely to be immense.

I thank Kip Thorne and Steinn Sigurdsson for help and advice in
writing this talk; I also thank Ramesh Narayan for allowing me to use
a preliminary draft of Ref.\ {\cite{sah:narayan}}.  I am indebted to
many people who helped me construct my radiation reaction code,
including (but not limited to) Sam Finn, Daniel Kennefick, Yuri Levin,
Amos Ori, Sterl Phinney, and ``The Capra Gang'': Lior Burko, Patrick
Brady, \'Eanna Flanagan, Eric Poisson, and Alan Wiseman.  This
research was support by NSF Grants AST-9731698 and AST-9618537, and
NASA Grants NAG5-6840 and NAG5-7034.

\end{document}